\documentclass[twocolumn,aps]{revtex4}
\usepackage{graphicx}
\usepackage{amsmath}
\usepackage{amsfonts}
\usepackage{amssymb}

\begin{document}

\title{\textbf{Self-Organized Criticality, Optimization and Biodiversity}}
\author{Roberto N. Onody}
\email{onody@if.sc.usp.br}
\author{Paulo A. de Castro}
\email{pac@if.sc.usp.br} \affiliation{Departamento de F\'{\i}sica
e Inform\'atica,
Instituto de F\'{\i}sica de S\~ao Carlos, \\
Universidade de S\~ao Paulo, C.P.369, 13560-970 S\~ao Carlos-SP,
Brazil}

\begin{abstract}
By driven to extinction species less or poorly adapted, the
Darwinian evolutionary theory is intrinsically an optimization
theory. We investigate two optimization algorithms with such
evolutionary characteristics: the Bak-Sneppen and the Extremal
Optimization. By comparing their mean fitness in the steady state
regime, we conclude that the Bak-Sneppen dynamics is more
efficient than the Extremal Optimization if the parameter $\tau$
is in the interval $[0,0.86]$. The determination of the spatial
correlation and the probability distribution of the avalanches
show that the Extremal Optimization dynamics does not lead the
system into a critical self-organized state. Trough a discrete
form of the Bak-Sneppen model we argument that biodiversity is an
essential requisite to preserve the self-organized criticality.


\end{abstract}
\maketitle

\section{Introduction}
 By all that we know nature evolves in a self-organized
 critical state \cite{1}. One of the most fundamental characteristics of a system in
 a self-organized critical state (SOC) is to exhibit a stationary state with
 a long-range power law decay of both spatial and temporal correlations \cite{2}.
 Power law is a very abundant behavior appearing either in natural phenomena
 such as the light emitted from quasars, the earthquakes intensities, the water
 level of the Nile river or as a direct result of human activities like the
 distribution of cities by size, the repetition of words in the Bible and
 in traffic jams.

 Self-organized critical systems evolve to the complex critical
 state without the interference of any external agent - there is no tuning parameter.
 The prototypical example of SOC is a pile of sand \cite{2}. Usually, the self-organized state
 is attained only after a very long period of transient. Last but not least, a minor
 change in the system can cause colossal instabilities called avalanches. Intermittent
 bursts of activity separating long periods of quiescence is called punctuated equilibrium.
 Gould and Eldredge conjectured that the biological evolution in our planet is under
 the auspices of this kind of mechanism \cite{3}.

 The evolution of the living beings is basically governed by the theory of natural selection.
 One model specially tailored to represent the co-evolutionary activities of the species
 is the Bak-Sneppen model (BS). In this model \cite{4}, each species occupies a site $i$ of a lattice
 and has associated a fitness value $\lambda_{i}$ between 0 and 1 ( randomly drawn from an uniform
 distribution). At each time step, the species with the smallest fitness as well as its nearest
 neighbors are selected to replace their fitness with new random numbers. In one dimension,
 after a long transient time, almost all species have fitness larger than the critical value
 0.67 \cite{4}.

 Recently, inspired by natural processes, some heuristic optimization techniques have been
 proposed: genetics algorithms \cite{5}, simulated annealing \cite{6} and extremal optimization
 \cite{7}. The latter, the extremal optimization method (EO), is the most efficient since it
 brings the system faster and closer to its ground state. In brief words, this method consists
 of the following rules: 1) a fitness $\lambda_{i}$ with values between 0 and 1 (randomly chosen
 from an uniform distribution) is associated with each site $i$ of a lattice with $N$ points;
 2) all the lattice sites are increasingly ranked according to their fitness (the site with the
 worst fitness is of rank 1); 3) a site of rank $k$ ($1 \le k \le N$) is selected with probability
 $ P(k)\sim k^{-\tau}$ ($\tau$ is an arbitrary real positive number) and its corresponding variable
 $\lambda_{i}$ is changed to $\lambda_{i}^{'}$ ;  4) repeat at step 2) as long as desired.

 We observe that, differently from what happens with the Bak-Sneppen dynamics, the EO dynamics has
 neither a co-evolutionary feature (the extinction of one species has no influence on its neighbors)
 nor has the exact (Darwinian) characteristic of the elimination of the worst adapted species. In
 this sense, we can say that the BS algorithm is a coarse grained description of the biological
 evolution adopted by nature while the EO algorithm represents an optimized dynamics created by
 man. In this paper, we compare the efficiencies (measured by their mean fitness in the steady state)
 of the EO and the BS dynamics. For the EO dynamics, we show that the spatial
 distribution is constant meanwhile the distribution of avalanches
 has an exponential decay. Using a discrete form of the BS model
 we argue that variability of species is an essential requisite to
 keep self-organized criticality.

\section{Simulations}
 To compare both dynamics, we simulated the BS and EO algorithms up to $1.1 \; 10^{9}$ runs on a one dimensional
 ring with N=4001
 sites. To guarantee that the stationary regime has been achieved, we discarded the first $1.0 \; 10^{8}$
 runs as the transient time. Time averages were then taken over the remaining steps. Figure 1 shows the
 average frequency of the fitness $\lambda$. Clearly, for $\tau = 0.05$ the EO behaves like a random
 walk having an almost uniform and constant fitness distribution. At $\tau = 0.5$ ($1.0$) the distribution is an
 increasing linear (exponential) function of $\lambda$. For the BS dynamics, however, the distribution
 has the form of a step function with a discontinuity at the critical point $\lambda_{c} \sim 0.67$. This
 critical point exists in all regular geometries or exponential networks \cite{8,9,10}, but not in
 scale-free networks \cite{11}. The presence of this critical point in the BS algorithm is the first
 sign that a critical self-organized state has been developed. For the EO there is no such a point.

\begin{figure}[htbp!]
\begin{center}
\includegraphics[width=8cm]{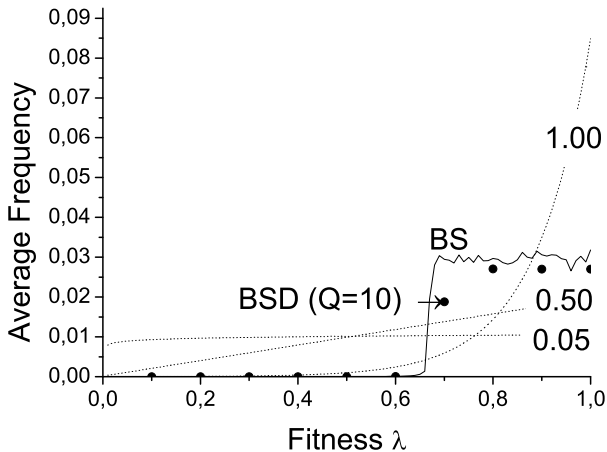}
\end{center}
\caption{The frequency of the fitness $\lambda$ averaged on time
after the stationary regime has been reached. The full line
corresponds to the BS algorithm and exhibits a discontinuity at
the critical point $\lambda_{c} \sim 0.67$. The dotted lines are
the EO algorithm with $\tau = 0.05$, $0.50$ and $1.00$. The points
represent the discrete BSD
algorithm with only 10 possible discrete fitness values (see the text).}%
\label{Figure1}%
\end{figure}

 To measure the algorithm's efficiencies, we plotted in Fig.2 the mean fitness obtained in the steady
 state regime. For the BS dynamics the mean fitness is $0.83$. This mean fitness corresponds to an EO with
 $\tau = 0.86$. At $\tau = 1.5$ the mean fitness of the EO dynamics is approximately $0.99$. This means
 that for higher values of $\tau$, the EO algorithm leads to an utopian society where only one and perfect
 species survives. This limit corresponds to the simplified toy model proposed by K. K. Yee \cite{12}
 in the context of law's evolution in the judicial system.
 For $\tau$ in the interval $[0, 0.86]$, the BS surpasses EO.

\begin{figure}[htbp!]
\begin{center}
\includegraphics[width=8cm]{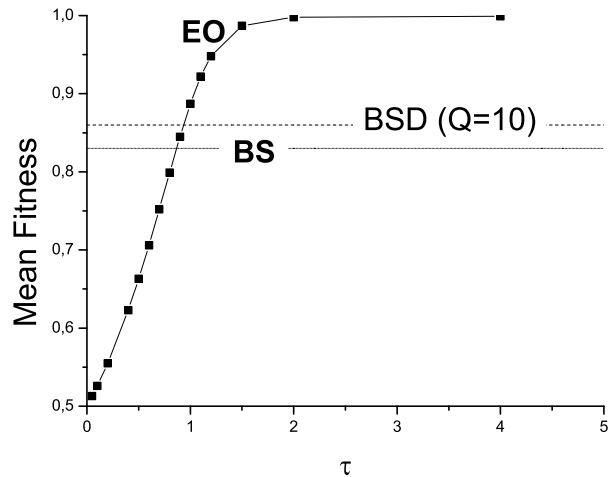}
\end{center}
\caption{The mean fitness of the EO dynamics for various values of $\tau$. The BS (BSD) algorithm has a
mean fitness value of $0.83$ ($0.86$) which corresponds to $\tau = 0.86$ ($0.95$) in the EO curve.}%
\label{Figure2}%
\end{figure}

 As we pointed out before, while BS is a co-evolutionary dynamics the EO dynamics is only evolutionary. The species
 in the EO do not interact. Comparison between co-evolutionary and evolutionary performances have already been
 done in the context of cellular automata \cite{13}. To investigate the main differences between the
 evolutionary (EO) and co-evolutionary (BS) dynamics, we studied their spatial correlation dependence.
 Let $D(x)$ be the
 probability distribution of the distance $x$ between two subsequent extinct (or mutated) species.
 From the Fig. 3, it is clear that the EO dynamics does not show a critical self-organized
 behavior. Instead of a power law, its spacial correlation is of infinite range - the probability distribution
 is constant no matter what is the distance between two subsequently modified species. For
 the BS dynamics, we find the well known
 power law dependence $D(x)\sim x^{-3.23 \pm 0.02}$ \cite{14}.

 \begin{figure}[htbp!]
\begin{center}
\includegraphics[width=8cm]{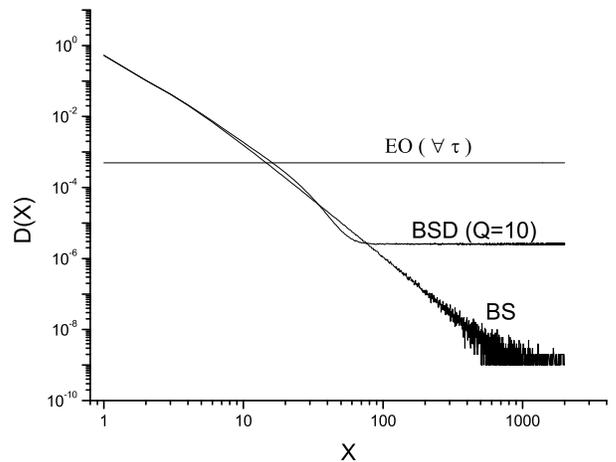}
\end{center}
\caption{The EO algorithm has a constant distribution probability
$D(x)$ which is independent of the
$\tau$ value. The BS algorithm shows a power law decay while the BSD has a mixed behavior.}%
\label{Figure3}%
\end{figure}

 Another important difference between the two dynamics is the complete absence of the punctuated equilibrium
 in the EO algorithm. One way to check out the existence of the punctuated
 equilibrium is to measure the probability distribution $P(A)$ of the avalanches with size $A$. The size $A$
 of an avalanche is defined as been the number of subsequent time steps with at least one fitness value below a
 critical threshold $\lambda_{c}$. This critical point doesn't exist for the EO (Fig. 1). For the BS algorithm
 the distribution decays as $P(A) \sim A^{-1.07 \pm 0.01}$ at $\lambda_{c} = 0.67$
 \cite{14}. In the EO algorithm, on the other hand, the decay is exponential with a
 characteristic avalanche size $A_{c}(\lambda_{c},\tau)$ depending
 on the choices made for $\lambda_{c}$ and $\tau$.

\section{Conclusions}

 We conclude that although the efficiency of the EO algorithm may exceed, under certain circumstances
 (if $\tau > 0.86$), that of the BS dynamics, it is accompanied by three undesirable characteristics:
 the spatial correlation between the species is constant, i. e., it is independent of their distances, there
 is an external free parameter $\tau$ to be adjusted by hand and the punctuated equilibrium mechanism is lost.
 The punctuated equilibrium seems to be a very productive form found by nature to innovate species
 without the intervening of climatic changes or meteors destruction.

 We have learned that the EO dynamics does not conduct the system to a critical self-organized state. However,
 we would like to point out that even the BS can loose its SOC characteristics and, amazingly, in a very easy and
 quick manner. Suppose that, instead a continuous and uniform fitness distribution in the interval $[0,1]$, only
 some discrete values are now possible. To simplify, assume that the fitness can only have $Q$ equally spaced
 values, i. e., $\lambda = m /Q$, with $m=1,2,...,Q$. Practically this means, that for some reason, the system's
 biodiversity has decreased. Due to the discreteness, there will be an enormous number of species carrying the
 same (worst) fitness value. Which species should then we choose? The simplest solution is to put all those species
 in a list and to draw one of them. We will call this dynamics as Bak-Sneppen with draw (BSD).
 In the
 Fig.1
 we plotted the case $Q=10$ and observe that, like in the EO dynamics, there is not a critical threshold $\lambda_{c}$.
 The mean fitness is $0.86$ (Fig.2), a value which is a little bit greater than that of the standard BS.
 The curve of the spatial
 probability distribution $D(x)$ (see Fig.3) is even more interesting. It shows that the BSD dynamics is of a
 mixed kind: it behaves like the BS for small distances and like the EO for large distances. So, the BSD dynamics
 does not retain the self-organized criticality characteristic. Just like in nature, biodiversity plays a fundamental
 role in the evolutionary theoretical models: without it self-organized criticality is not possible. For higher plants
 and animals the conventional explanations of biodiversity are habitat heterogeneity, predation pressure and niche
 differentiation. For microscopic organisms, however, the high biodiversity found (even in uniform environments) is not
 completely understood and it is called "the paradox of the plankton". Theoretically, such difficulties can be
 surmounted by incorporating a noise $\eta$ into the fitness $\lambda = m /Q + (1-2r)\eta$ (where $r$ is a random number
 in the interval $[0,1]$ and generated from a uniform distribution). Even for $\eta$ as small as $10^{-12}$ the SOC
 characteristic is preserved \cite{15}. The noise can be interpreted as the presence of sub-species.

This work was supported by CNPq (Conselho Nacional de
Desenvolvimento Cient\'{\i}fico e Tecnol\'ogico)
  and FAPESP (Funda\c c\~ ao de Amparo \`a Pesquisa do Estado de S\~ao Paulo).

\end{document}